\documentclass[preprint,showpacs,preprintnumbers,amsmath,amssymb]{revtex4}

\usepackage{graphicx}
\usepackage{dcolumn}
\usepackage{bm}

\begin{document}


\title{An Optical Clock Based on Coherent Population Trapping of Alkaline-earth Ions}

\author{Z. H. Lu}
\author{L. J. Wang}%
\email{Lijun.Wang@mpl.mpg.de}
\affiliation{%
Max-Planck Institute for the Science of Light\\
Guenther-Scharowsky-Str. 1, Building 24\\
91058 Erlangen, Germany
}%

\date{\today}

\begin{abstract}
An ultra-stable optical clock based on coherent population trapping effect of alkaline-earth ions, such as Ca$^+$, Sr$^+$, Ba$^+$, is analyzed here. The proposed transitions use the odd isotopes, so that the frequency shift is insensitive to the applied magnetic field. To enhance the signal, a large number of ions are trapped in a linear Paul trap, and laser cooled to crystallize such that the ions are in the Lamb-Dicke regime to avoid first order Doppler shift. Other relevant frequency shifts are also analyzed. Further techniques to improve the stability of the clock signal are discussed. 
\end{abstract}

\pacs{06.30.Ft, 37.10.Ty, 32.70.Jz}
\maketitle

Optical clocks, or atomic clocks using optical transitions, have much higher line Q comparing to microwave atomic clocks. As a consequence, optical clocks potentially can have better stability and accuracy than today's microwave atomic clocks. Many groups are working on optical clocks with the goal of realizing improved frequency standards and testing fundamental physics. Commonly, there are two ways of making an optical clock. One is based on a single, tightly trapped ion with very small systematics uncertainties. The disadvantage of this method is the intrinsically smaller signal strength. Consequently, a long sampling time will be required to realize the potential of this type of optical clocks. This problem can be alleviated by confining many neutral atoms in an optical lattice trap with much improved signal strength \cite{Katori}. In this case, the relevant atomic levels shift equally (via AC Stark shift) in the optical lattice, at the so-called ``magic wavelength.'' Both approaches, so far, operate in pulsed interrogation mode. Pulsed interrogation, such as ``shelving,'' has a low duty cycle that relies heavily on the local oscillator's free-running stability. In addition, there could be complicated aliasing problems such as the Dick effect \cite{Dick}.

In this Letter, we analyze a third approach, namely, an optical clock that operates continuously using the coherent optical trapping (CPT) transitions of a large, laser-cooled, trapped ion crystal. We consider a cloud of ions trapped in a large, linear ion trap. In this case, the ions will be cooled to sufficiently low temperature such that an ion crystal is formed. At such low temperatures ($\sim 1$ mK), the ions' secular motion amplitude is less than half of the laser wavelength, and we are in the so-called Lamb-Dicke regime \cite{Dicke}, resulting in a vanishing first-order Doppler shift. This situation is identical to the optical lattice, where an artificial optical potential is applied to confine the atoms. In the case analyzed here, simple laser cooling will crystallize and confine the ions, without the additional lattice laser. Under such conditions, it is possible to construct an optical clock on the sharp dark lines of CPT with vanishingly small systematics.

CPT, a phenomenon that is accomplished by two coherent lasers interacting in a $\Lambda$ scheme \cite{Alzetta, Arimondo}, is generally observed in alkali atoms. There are many applications with CPT, including microwave atomic frequency standards \cite{Vanier} and magnetometry \cite{Scully, Nagel}. Narrow linewidths below $50$ Hz have been realized in buffer-gas-filled vapor cells \cite{Brandt, Merimaa}. This corresponds to relative clock frequency stability as low as $1.3\times10^{-12}$ for $1$ s integration time. Even better stability can be expected with an optical CPT. In this Letter, we propose an optical frequency standard based on optical CPT, as shown in Fig. 1(a). The proposed scheme is applicable to alkaline-earth ions, such as Ca$^+$, Sr$^+$, and Ba$^+$. But as an example, we will focus on Ba$^+$. In the proposed scheme, two phase-locked lasers at $493$ nm (green laser) and $650$ nm (red laser) will interact with crystallized Barium ions, and spectrally narrow light at the exact clock frequency can be generated through frequency mixing as the frequency standard.

More specifically, odd isotope $^{137}$Ba$^+$ will be chosen, and the three energy levels involved are $\left|6S_{1/2}, F=2\right\rangle$, $\left|6P_{1/2}, F=1\right\rangle$, and $\left|5D_{3/2}, F=0\right\rangle$, as shown in Fig. 1(b). It is possible then to use circular polarized light to establish coherence between $\left|6S_{1/2}, F=2, m_F=0\right\rangle$, and $\left|5D_{3/2}, F=0, m_F=0\right\rangle$. The transition between $m_F=0\leftrightarrow m_F=0$ is chosen so that they are independent of the applied magnetic field in first order, a desirable property in frequency standards applications.

To understand the full dynamics of the proposed system under magnetic field, a twenty-level density matrix would have to be used. But the basic phenomenon that we are interested in can be understood by using a three-level model. Under the rotating wave approximation, the time evolution of the three-level atomic system coupled to the two light fields is governed by the Liouville's equation,
\begin{equation}
\frac{d\rho}{dt}=-\frac{i}{\hbar}\left[H,\rho\right]+\textsl{L}(\rho),
\label{bloch}
\end{equation}
where $\rho$ is the atomic density matrix, and we have
\begin{equation}
\text{Tr}[\rho]=\rho_{11}+\rho_{22}+\rho_{33}=1.
\end{equation}
The system Hamiltonian is defined as
\begin{equation}
H=\hbar\left(\begin{array}{ccc} \Delta_g & \Omega_{12}/2 & 0 \\
                                \Omega_{12}/2 & 0 & \Omega_{23}/2 \\
                                0 & \Omega_{23}/2 & \Delta_r
              \end{array} \right),
\end{equation}
where $\Delta_g=\omega_g-\omega_{12}$ and $\Delta_r=\omega_r-\omega_{23}$ are the detunings between the green ($\left|1\right\rangle\rightarrow \left|2\right\rangle$) and red ($\left|2\right\rangle\rightarrow \left|3\right\rangle$) laser frequencies, $\omega_g$ and $\omega_r$, and the atomic resonance frequencies, $\omega_{12}$ and $\omega_{23}$, respectively. $\Omega_{12}$ and $\Omega_{23}$ are the Rabi frequencies associated with the couplings of the green laser and the red laser to atomic transitions $\left|1\right\rangle\rightarrow \left|2\right\rangle$ and $\left|2\right\rangle\rightarrow \left|3\right\rangle$, respectively. The term $\textsl{L}(\rho)$ includes atomic relaxation terms and laser linewidths effect, and can be written as
\begin{equation}
\textsl{L}(\rho)=-\frac{1}{2}\sum_{n}(C^{\dag}_{n}C_n\rho+\rho C^{\dag}_{n}C_n-2C_n\rho C^{\dag}_{n}).
\label{L}
\end{equation}
Since $5D_{3/2}$ state has a very long lifetime ($80$ s), we can neglect its decay to $6S_{1/2}$ state. The decay from $6P_{1/2}$ state to $6S_{1/2}$ state is described by
\begin{equation}
C_{21}=\sqrt{\Gamma_{21}}\left|1\right\rangle\left\langle 2\right|,
\label{c1}
\end{equation}
where $\Gamma_{21}=2\pi\times 15.1$ MHz. Similarly, the decay from $6P_{1/2}$ state to $5D_{3/2}$ state is described by
\begin{equation}
C_{23}=\sqrt{\Gamma_{23}}\left|3\right\rangle\left\langle 2\right|,
\label{c2}
\end{equation}
where $\Gamma_{23}=2\pi\times 5.3$ MHz. The effect of laser linewidths can be introduced by
\begin{eqnarray}
C_g&=&\sqrt{2\Gamma_{g}}\left|1\right\rangle\left\langle 1\right|, \label{g}\\
C_r&=&\sqrt{2\Gamma_{r}}\left|3\right\rangle\left\langle 3\right|. \label{r}
\end{eqnarray}
Here, $\Gamma_{g}$ and $\Gamma_{r}$ denote the laser linewidths of the green and red lasers, respectively. Substituting Eq. \eqref{c1}, Eq. \eqref{c2}, Eq. \eqref{g}, and Eq. \eqref{r} to Eq. \eqref{L}, we obtain the Liouvillian matrix $\textsl{L}(\rho)$,
\begin{equation}
\textsl{L}(\rho)=\left(\begin{array}{ccc} \Gamma_{21}\rho_{22} & -(\Gamma_{21}+\Gamma_{23}+2\Gamma_{g})\rho_{12}/2 & -(\Gamma_g+\Gamma_r)\rho_{13} \\
                                -(\Gamma_{21}+\Gamma_{23}+2\Gamma_{g})\rho_{21}/2 & -(\Gamma_{21}+\Gamma_{23})\rho_{22} & -(\Gamma_{21}+\Gamma_{23}+2\Gamma_{r})\rho_{23}/2 \\
                                -(\Gamma_g+\Gamma_r)\rho_{31} & -(\Gamma_{21}+\Gamma_{23}+2\Gamma_{r})\rho_{32}/2 & \Gamma_{23}\rho_{22}
              \end{array} \right).
\end{equation}

Under stationary condition where $\dot{\rho}=0$, Eq. \eqref{bloch} can be solved to obtain the resonance fluorescence excitation spectrum under different detunings. When $\Delta_g=\Delta_r$, a CPT dark line will appear, as shown in \cite{Janik,Siemers,Stalgies}. The linewidth of the dark line depends on laser linewidths and laser power. If we neglect the effect of laser linewidths, then the CPT linewidth at optical resonance is
\begin{equation}
\Delta\nu=\frac{\gamma}{\pi\sqrt{1+S}}\sqrt{(1+3S+\frac{\Gamma}{\gamma}S^*)(1+\frac{\Gamma}{\gamma}S^*)}\ .
\end{equation}
Here, $\gamma$ is the decay rate of the $5D_{3/2}$ state, $\Gamma=\Gamma_{21}+\Gamma_{23}$, and $S$ and $S^*$ are two saturation parameters,
\begin{eqnarray}
S&=&\frac{\Omega^2_{12}\Omega^2_{23}}{\Omega^2_{12}+\Omega^2_{23}}\frac{4}{\Gamma^2}\ , \\
S^*&=&\frac{\Omega^2_{12}+\Omega^2_{23}}{4\Gamma^2}\ .
\end{eqnarray}
At zero laser powers, the CPT linewidth is limited by the decay rate of the $5D_{3/2}$ state at $2$ mHz. More realistic value of $\sim 10$ Hz can be achieved with reasonable laser powers. As an example, using $\Omega_{12}/2\pi=\Omega_{23}/2\pi=10$ kHz, the calculated linewidth is $5$ Hz, as shown in Fig.~2(a). Experimentally, the peak of this dip can be locked by dithering the driving frequency of an AOM that is used to scan the repump laser frequency. Assuming $10^6$ ions with $1\%$ collection efficiency, we can achieve a PMT counting rate of around $10$ kHz, from which we can estimate that the proposed frequency standard has stability of better than $10^{-16}$ in $1$ s in the shot-noise-limited photon counting situation. 

At this moment, it is worthwhile to estimate all systematic frequency shifts of the proposed clock transitions. In second-order perturbation theory, the light shift of state $\left|k,m\right\rangle$ under light beam of frequency $\omega$, electric field strength $E$, and polarization $\hat{\epsilon}$ can be calculated as
\begin{equation}
\Delta\nu_{k,m}=\frac{1}{4}\frac{E^2}{h}\sum_{k',m',\pm \omega} \frac{\left|\left\langle k,m|\hat{\epsilon}\cdot e\vec{r}|k',m'\right\rangle\right|^2}{W_k-W_{k'}\pm\hbar\omega}.
\end{equation}
Here, $k$ stands for atomic quantum number $nJF$, and $W_k$ is the unperturbed energy of level $k$. The sums extend over continuum states. In our case, for $\left|6S_{1/2}, F=2\right\rangle$ state, the main light shift contributions come from the near-resonant transition light of $493$ nm connecting $\left|6S_{1/2}, F=2\right\rangle$ to $\left|6P_{1/2}, F=1\right\rangle$ and $\left|6P_{1/2}, F=2\right\rangle$ states; for $\left|5D_{3/2}, F=0\right\rangle$ state, the main light shift contributions come from the near-resonant transition light of $650$ nm connecting $\left|5D_{3/2}, F=0\right\rangle$ to $\left|6P_{1/2}, F=1\right\rangle$ state. As a consequence, we can estimate the light shift of the $\left|6S_{1/2}, F=2\right\rangle$ state under right-circular polarized light as
\begin{eqnarray}
\Delta\nu_{6S}&=&\frac{4\pi\Omega_{12}^2}{3\epsilon_0\lambda_g^3\Gamma_{21}}[\frac{\left|\left\langle 6S_{1/2},F=2,m=0|\mu_1^1|6P_{1/2},F'=1,m'=1\right\rangle\right|^2}{\hbar\Delta_g} \nonumber \\ 
&&+\frac{\left|\left\langle 6S_{1/2},F=2,m=0|\mu_1^1|6P_{1/2},F'=2,m'=1\right\rangle\right|^2}{\hbar(\Delta_g-\delta_P)}] ,
\label{6s}
\end{eqnarray}
where $\delta_P/2\pi=1488$ MHz is the $6P_{1/2}$ state energy hyperfine splitting. Here, we have used the fact that
\begin{eqnarray}
I_g&=&\frac{1}{2}c\epsilon_0 E_g^2 \nonumber \\
&=&\frac{8\pi}{3}\frac{ch}{\lambda_g^3\Gamma_{21}}\Omega_{12}^2.
\end{eqnarray}
The two matrix elements in Eq.~\eqref{6s} can be calculated as
\begin{eqnarray}
\left|\left\langle 6S_{1/2},F=2,m=0|\mu_1^1|6P_{1/2},F'=1,m'=1\right\rangle\right|&=&\frac{1}{2\sqrt{6}}\mu_g, \\
\left|\left\langle 6S_{1/2},F=2,m=0|\mu_1^1|6P_{1/2},F'=2,m'=1\right\rangle\right|&=&\frac{1}{2\sqrt{2}}\mu_g,
\end{eqnarray}
where the reduced matrix element $\mu_g=\left|\left\langle 6S_{1/2}||\mu^1||6P_{1/2}\right\rangle\right|=8.43$ Debye. Similarly, the light shift of the $\left|5D_{3/2}, F=0\right\rangle$ state under right-circular polarized light is
\begin{equation}
\Delta\nu_{5D}=\frac{4\pi\Omega_{23}^2}{3\epsilon_0\lambda_r^3\Gamma_{23}}\frac{\left|\left\langle 5D_{3/2},F=0,m=0|\mu_1^1|6P_{1/2},F'=1,m'=1\right\rangle\right|^2}{\hbar\Delta_r}, 
\label{5d}
\end{equation}
where
\begin{equation}
\left|\left\langle 5D_{3/2},F=0,m=0|\mu_1^1|6P_{1/2},F'=1,m'=1\right\rangle\right|=\frac{1}{2\sqrt{3}}\mu_r,
\end{equation}
with the reduced matrix element $\mu_r=\left|\left\langle 5D_{3/2}||\mu^1||6P_{1/2}\right\rangle\right|=7.64$ Debye. By adjusting the relative intensities of the green and red lasers, it is possible to suppress the light shift to a negligible level. For example, using the parameters of Fig. 2(b), with $\Omega_{12}/2\pi=14$ kHz and $\Omega_{23}/2\pi=10$ kHz, the total light shift can be calculated as
\begin{equation}
\Delta\nu_{LS}=\Delta\nu_{5D}-\Delta\nu_{6S}=-4.8\ \text{mHz}.
\end{equation}
This value can be further decreased if we choose greater detunings. With the parameters of Fig. 2(b), the transition linewidth is increased, but the signal strength is similarly increased, so the resulting clock stability is still effectively the same.

For the proposed clock transition, the first-order Zeeman shift is zero, and the quadrupole DC-Stark shift due to stray electric field gradients is also zero in first-order. The accuracy of the proposed frequency standard is primarily limited by second-order Doppler shift, second-order Zeeman shift, blackbody Stark shift, and quadrupole AC-Stark shift. 

The second-order Doppler shift can be estimated as
\begin{equation}
\frac{\Delta\nu_D}{\nu}=-\frac{\left\langle v^2\right\rangle}{2c^2}=-(2\left\langle n\right\rangle+1)\frac{\hbar\omega_s}{4Mc^2}.
\end{equation}
Assuming a secular frequency $\omega_s/2\pi=0.1$ MHz, and the mean occupation number $\left\langle n\right\rangle=100$, the second-order Doppler shift can be estimated to be $-1.65\times10^{-19}$. Using Breit-Rabi formula \cite{Ramsey}, the second-order Zeeman shift can be calculated as
\begin{equation}
\frac{\Delta\nu_{Zeeman}}{\nu}=\frac{\Delta_{5D,F=0,m=0}-\Delta_{6S,F=2,m=0}}{\nu}=(-7.55\times10^{-11}/G^2)B^2.
\end{equation}
With an applied $1$ mG magnetic field and a $0.1$ mG stability, the second-order Zeeman shift is $-7.55\times10^{-17}$ with a relative uncertainty of $7.55\times10^{-18}$. The blackbody Stark shift can be calculated as
\begin{equation}
\frac{\Delta\nu_{BB}}{\nu}=\frac{\Delta_{BB,5D}-\Delta_{BB,6S}}{\nu}=6.37\times10^{-17}(V/cm)^{-2}\left\langle E^2\right\rangle.
\end{equation}
Here the time averaged electric field is \cite{Itano}
\begin{equation}
\left\langle E^2\right\rangle=(8.319\ V/cm)^2\left(\frac{T}{300\ K}\right)^4.
\end{equation}
At $300$ K room temperature, we can estimate the blackbody Stark shift to be $-5.1\times10^{-15}$. With a temperature stability of $1\ ^o$C, we estimate an uncertainty of the shift at $6.8\times10^{-17}$. The $6$S$_{1/2}$ quadrupole AC-Stark shift is zero, while $5$D$_{3/2}$ quadrupole AC-Stark shift is estimated to be $0.2$ mHz if we assume $1$ mm trap size and $200$ V rf voltage, so the final quadrupole AC-Stark shift is
\begin{equation}
\frac{\Delta\nu_{Stark}}{\nu}=\frac{\Delta_{Stark,5D}}{\nu}=1.37\times10^{-18}.
\end{equation}
Hence, we can see that the dominant systematics is due to blackbody Stark shift. However, its influence on clock stability can be eliminated by better control of the surrounding temperature variation.

In order to realize the proposed frequency standard, there are several practical issues that need to be dealt with. First, the CPT process connect $\left|6S_{1/2},F=2\right\rangle$, $\left|6P_{1/2},F=1\right\rangle$, and $\left|5D_{3/2},F=0\right\rangle$ states, but $\left|6P_{1/2},F=1\right\rangle$ can also decay to $\left|6S_{1/2},F=1\right\rangle$, $\left|5D_{3/2},F=1\right\rangle$, and $\left|5D_{3/2},F=2\right\rangle$ states. This problem can be solved by using another set of auxiliary cooling and repumping lasers to close the pumping cycle. Second, since we use circular polarized light, atoms are optically pumped to a Zeeman sublevel of $\left|6S_{1/2},F=2\right\rangle$ that is not involved in the CPT process. This effect has consequences on the observed signal amplitude. To fully account for this effect, an effective four level model would have to be used \cite{Vanier2}. This effect of optical pumping can be reduced by using a double lambda scheme \cite{Windholz, Jau}. Finally, in order for the proposed CPT scheme to work, the laser linewidths need to be stabilized to the Hz level. This is readily achievable with current development of ultra-stable high finesse reference cavities \cite{Stoehr,Ludlow,Alnis,Liu}. Also, precise relative phase coherence between the green and red lasers are required. This can be achieved by locking a frequency comb to one of the laser, serving as the master laser, and phase-lock another laser to the frequency comb with a tunable frequency offset.

To further improve the stability of the frequency standard, we need to further decrease the CPT linewidth and increase the signal strength. This is difficult to do in continuous CPT due to optical power saturation. An alternative method to overcome this limit is to generate optical CPT pulses to observe Ramsey fringes \cite{Thomas1,Thomas2}. With this method, the signal strength can be increased greatly, while the Ramsey fringe widths are dependent only on the transit time $T$ between the interaction zones as $1/2T$. This time domain Ramsey method has been demonstrated in microwave CPT \cite{Zanon,Zanon2,Zanon3}, and we believe it will also work for optical CPT.

In conclusion, we propose and analyze in details an ultra-stable optical frequency standard based on optical CPT. The proposed frequency standard can work in the continuous-wave mode and has a frequency stability at the $10^{-16}$ level, with very favorable systematics uncertainties. Even better stability can be achieved with Ramsey CPT methods.

ZHL acknowledges fellowship support from Alexander von Humboldt Foundation.


\newpage

\begin{figure}[tfb]
\centerline{\includegraphics[width=14cm]{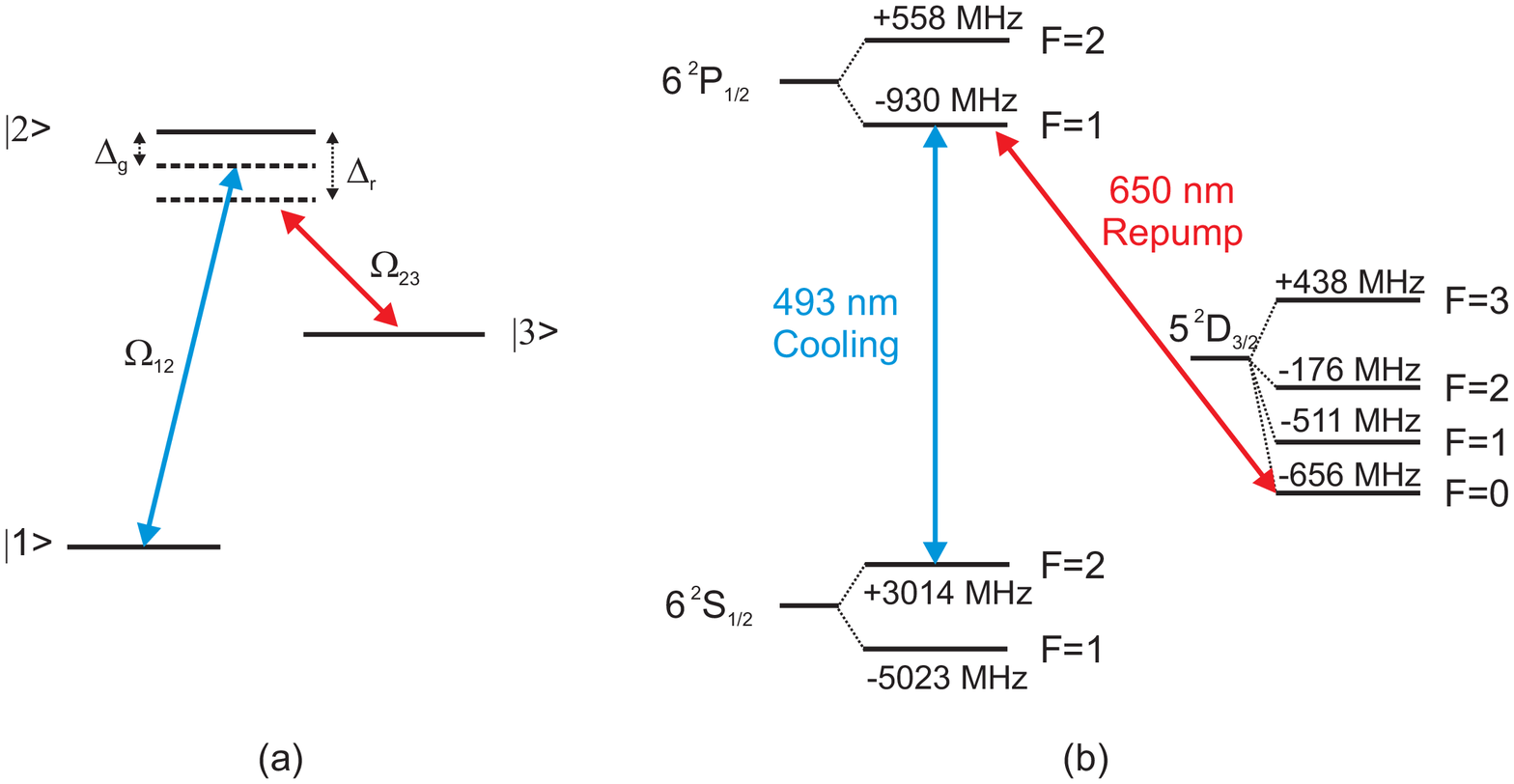}}
\caption{(Color online) (a) Energy level structure and optical couplings of the three-level atomic system for making a CPT frequency standard. (b) Energy level structure of odd isotope Barium ion.}
\label{fig1}
\end{figure}

\begin{figure}[tfb]
\centerline{\includegraphics[width=16cm]{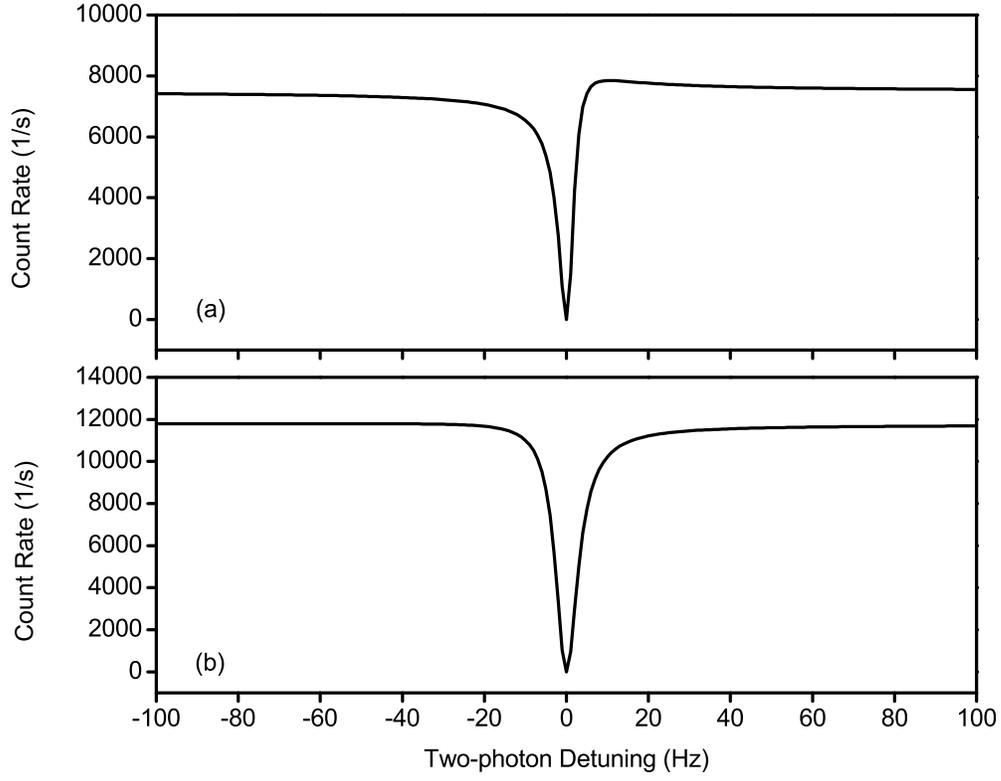}}
\caption{Resonance fluorescence excitation spectrum with different red laser detuning. Assuming $10^6$ ions and $1\%$ collection efficiency. (a) $\Delta_g/2\pi=-20$ MHz, $\Omega_{12}/2\pi=\Omega_{23}/2\pi=10$ kHz, $\Gamma_g=\Gamma_r=0$. (b) $\Delta_g/2\pi=-20$ MHz, $\Omega_{12}/2\pi=14$ kHz, $\Omega_{23}/2\pi=10$ kHz, $\Gamma_g=\Gamma_r=0$.}
\label{fig2}
\end{figure}

\end{document}